\documentclass[10pt]{iopart}

\usepackage{graphicx}
\usepackage{color}
\usepackage{cite}

\begin{document}

\topical{Multiple photodetachment of atomic anions via single and double core-hole creation}

\author{S~Schippers$^1$, A~Perry-Sassmannshausen$^1$, T~Buhr$^1$, M~Martins$^2$, S~Fritzsche$^{3,4}$, and A~M\"uller$^5$}
\address{$^1$I. Physikalisches Institut, Justus-Liebig-Universit\"at Gie\ss{}en, Heinrich-Buff-Ring 16, 35392 Giessen, Germany}
\address{$^2$Institut f\"{u}r Experimentalphysik, Universit\"{a}t Hamburg, Luruper Chaussee 149, 22761 Hamburg, Germany}
\address{$^3$Helmholtz-Institut Jena, Fr{\"o}belstieg 3, 07743 Jena, Germany}
\address{$^4$Theoretisch-Physikalisches Institut, Friedrich-Schiller-Universit\"{a}t Jena, 07743 Jena, Germany}
\address{$^5$Institut f\"{u}r Atom- und Molek\"{u}lphysik, Justus-Liebig-Universit\"{a}t Gie{\ss}en, Leihgesterner Weg 217, 35392 Giessen, Germany}

\ead{stefan.schippers@physik.uni-giessen.de}
\vspace{10pt}
\begin{indented}
\item[]\today
\end{indented}

\begin{abstract}
We review the recent experimental and theoretical progress in $K$-shell detachment studies of atomic anions. On the experimental side, this field has largely benefitted from technical advances at 3rd generation synchrotron radiation sources. For multiple detachment of C$^-$, O$^-$, and F$^-$ ions, recent results were obtained at the photon-ion merged-beams setup PIPE which is a permanent end station at beamline P04 of the PETRA\,III synchrotron light source in Hamburg. In addition to a much increased photon flux as compared to what was available previously, the PIPE setup has an extraordinary detection sensitivity for heavy charged reaction products that allows one to study detachment processes with extremely low cross sections in the kilobarn range, e.g., for processes involving the simultaneous creation of two core-holes by a single photon. The experimental findings pose new challenges for state-of-the-art atomic theory and require calculations combining photoexcitation (ionization) with decay cascade processes that follow after initial core-hole production.
\end{abstract}

%
%
\submitto{\JPB}
%
%
\ioptwocol

\section{\label{intro}Introduction}

Negative ions (anions \cite{Andersen2004b}) play an important role in low-temperature plasmas such as the upper atmosphere or the interstellar medium \cite{Millar2017} and also in technical applications. For example, in the context of antihydrogen production, it has been proposed to use an ensemble of laser-cooled anions as a coolant for antiprotons \cite{Kellerbauer2006}.

Atomic anions are highly correlated systems where the extra electron is weakly bound to an overall neutral charge distribution. For similar reasons, the number of excited states of atomic anions is quite limited, and some atomic species such as nitrogen do not form anions at all. A thorough treatment of the correlation effects in negative atomic ions still poses a formidable challenge to atomic theory, and this is even enhanced for the creation of core-holes, since the valence electrons are then subject to relaxation mediated by strong many-electron effects \cite{Gorczyca2004a,Schippers2016a}. On the experimental side, core holes can be created selectively by exciting or ionizing an inner-shell electron with a photon. For light ions, the core-hole states then subsequently autoionize which results in a net (multiple) ionization by a single photon. For negative ions, the entire process is termed (multiple) photodetachment. Corresponding experiments require energetic photons from a synchrotron radiation source.

Experimental work on inner-shell photodetachment of negative ions is scarce. Prior to the start of this project, photodetachment via the creation of a single $K$-hole has been experimentally studied only for a limited number of light anions up to O$^-$ \cite{Kjeldsen2001a,Berrah2001,Berrah2007a,Gibson2003a,Walter2006a,Gibson2012}. In these studies only double detachment has been considered, and this only for rather narrow energy ranges around the respective $K$-shell detachment thresholds. Further work on inner-shell photodetachment addressed the $L$, $M$ and $N$ shells of heavier atomic anions (see, e.g., \cite{Covington2001b,Kjeldsen2004b,Bilodeau2005a,Bilodeau2009,Dumitriu2017}). All these measurements were carried out mostly more than a decade ago using the photon-ion merged-beams technique \cite{Kjeldsen2006a,Schippers2016} (see below) at 3rd generation synchrotron light sources. At that time, the photon flux was considerably lower than what is currently available.

Here, we summarize the significant experimental progress that was recently achieved on multiple photodetachment via $K$-shell excitation and ionisation of C$^-$ \cite{Perry-Sassmannshausen2020}, O$^-$ \cite{Schippers2016a}, and F$^-$ \cite{Mueller2018b} ions at a high photon-flux beamline of a modern 3rd generation synchrotron in combination with intense ion-beams as well as sensitive and selective ion-detection techniques. This topical review is organized as follows. Section \ref{sec:method} provides a brief description of the photon-ion merged-beams technique. Experimental findings on inner-shell photodetachment of atomic anions are presented and discussed in section~\ref{sec:results}. Corresponding theoretical developments are presented in section~\ref{sec:theo}. Section~\ref{sec:conc} provides conclusions and a brief outlook.

\section{\label{sec:method}The photon-ion merged-beams technique}

\begin{figure*}
\includegraphics[width=\textwidth]{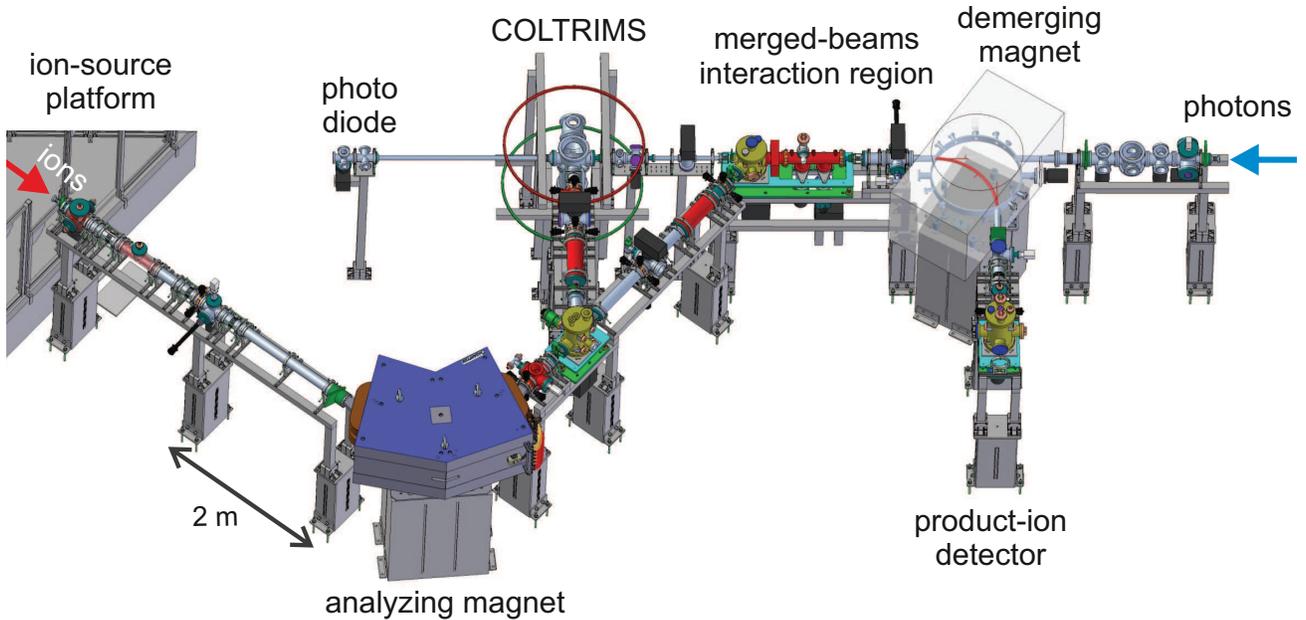}
\caption{\label{fig:PIPE} Sketch of the photon-ion merged-beams setup PIPE \cite{Schippers2014} which is a permanently installed end station at the photon beamline P04 \cite{Viefhaus2013} of the synchrotron light source PETRA III in Hamburg. The photon beam enters the setup from the right (blue arrow) and is parallel to the floor at a nominal height of 2.07~m. It is stopped by a calibrated photodiode (PD) which continuously monitors the absolute photon flux. The ion beam enters from the left (red arrow). It is generated with an ion source that is mounted on a separate platform (not fully shown). The analysing magnet (AM) provides mass/charge selection of ions for the further ion-beam transport. Spherical deflectors (SD) can be used to direct the ions either into the crossed-beams (CB) interaction point or into the merged-beams (MB) collinear beam overlap region. The demerging magnet (DM) deflects primary and product ions out of the photon-beam axis and directs product ions into the single-particle detector (SPD). The ion current can be measured at various places along the ion beamline by inserting Faraday cups (FC) into the ion beam. One FC is mounted inside the DM such that product ions which are deflected by $90^\circ$ can pass towards the SPD, while ions in different charge states or with different kinetic energies are collected in this FC. The MB is equipped with scanning slits for beam-profile measurements. The latest addition to PIPE is an ion trap (IT) equipped with a time-of-flight mass spectrometer. Adapted from \cite{Schippers2014}, copyright 2014 by IOP.}
\end{figure*}

Experiments with ionic targets are challenging because of the low target densities which are usually orders of magnitude smaller than the typical densities of neutral gas targets. The photon-ion merged-beams method (see \cite{Schippers2016} for a recent introductory review) compensates the low target density by providing an elongated interaction region of typical $\sim$1~m length where the photon beam and the ion beam move coaxially. In addition, heavy charged photo products can be collected and detected with nearly 100\% efficiency, since they move with keV energies and can be easily separated from the primary beam by electric or magnetic fields. Nevertheless the signal rates from such an arrangement are still quite small, such that meaningful experiments with UV and x-ray photons could only be carried out after the advent of 2nd-generation synchrotron light sources. Pioneering work was carried out at the Daresbury Synchrotron Radiation Source \cite{Lyon1987}. Since then, the technique spread to other synchrotron radiation sources, e.g., ASTRID \cite{Kjeldsen1999b}, SPring-8 \cite{Oura2001}, ALS \cite{Covington2002}, and SOLEIL \cite{Gharaibeh2011a,Bizau2016a}. Because of their rather large size (heavy magnets and ion sources) these ion-beam setups were realized as permanent installations and, hence, the available photon-energy range depends on the chosen photon beam line.

The latest development in this field is the \underline{P}hoton-\underline{I}on Spectrometer at \underline{PE}TRA\,III (PIPE) \cite{Schippers2014,Mueller2017} (figure~\ref{fig:PIPE}) which has been set up by a consortium of German university groups at the photon beamline P04 \cite{Viefhaus2013} of the PETRA\,III synchrotron operated by DESY in Hamburg, Germany. The PIPE setup is unique with respect to the available photon flux and photon energy range. PIPE is the only photon-ion merged-beams setup where photon energies higher than 1000 eV, i.e., up to nearly 3000~eV are available.

The PIPE setup features an ion-source platform where user-supplied ion sources can be easily installed such that the optimal source for a given ion species can be operated. The most recent addition is a Cs-sputter ion source for the production of intense beams of negatively charged ions which has been used for the experiment with C$^-$ ions discussed below \cite{Perry-Sassmannshausen2020}. This source had not been available for the previous experiments with O$^-$ \cite{Schippers2016a} and F$^-$ \cite{Mueller2018b} ions where an electron-cyclotron resonance (ECR) ion source was used with negative ion currents that were an order of magnitude lower than the C$^-$ ion current from the Cs-sputter ion source.

A salient feature of the photon-ion merged-beams method is its ability to provide absolute cross-section data which are important, e.g., for applications in plasma physics and astrophysics. To this end, the mutual spatial overlap between the ion beam and the photon beam has to be determined. At PIPE, three sets of slit scanners serve this purpose. The entire procedure gives rise to a systematic uncertainty of the resulting absolute cross section scale that amounts to typically $\pm15\%$ \cite{Schippers2014}.

The photon energy scale of the PETRA\,III photon beamline P04 can be calibrated by comparing absorption measurements in gases with previously established reference standards from the literature \cite{Mueller2017}. The systematic uncertainty of the photon-energy scale that results from this procedure is typically about $\pm200$~meV. This rather large error is directly related to the uncertainties of the reference standards in the XUV photon-energy range. Based on measurements at PIPE, it has recently been suggested that photoionisation resonances of few-electron ions can potentially serve as much more accurate reference standards \cite{Mueller2018c}. Corresponding activities are ongoing.

The PIPE setup has lead to a breakthrough in experimental sensitivity, which allows for much more detailed studies of inner-shell ionisation/detachment of ionic targets than was possible before \cite{Mueller2015a,Mueller2015,Mueller2018}.  The results from the first five years of operation of the PIPE setup have been summarized in \cite{Schippers2020}. Apart from inner-shell photodetachment and photoionisation of negatively and positively charged atomic ions also photo processes involving molecular ions, and positively and negatively charged endohedral fullerene ions have been investigated (see \cite{Bari2019,Schubert2019,Mueller2019} for the most recent corresponding results).

\section{\label{sec:results}Experimental findings}

\begin{figure*}
\includegraphics[width=\textwidth]{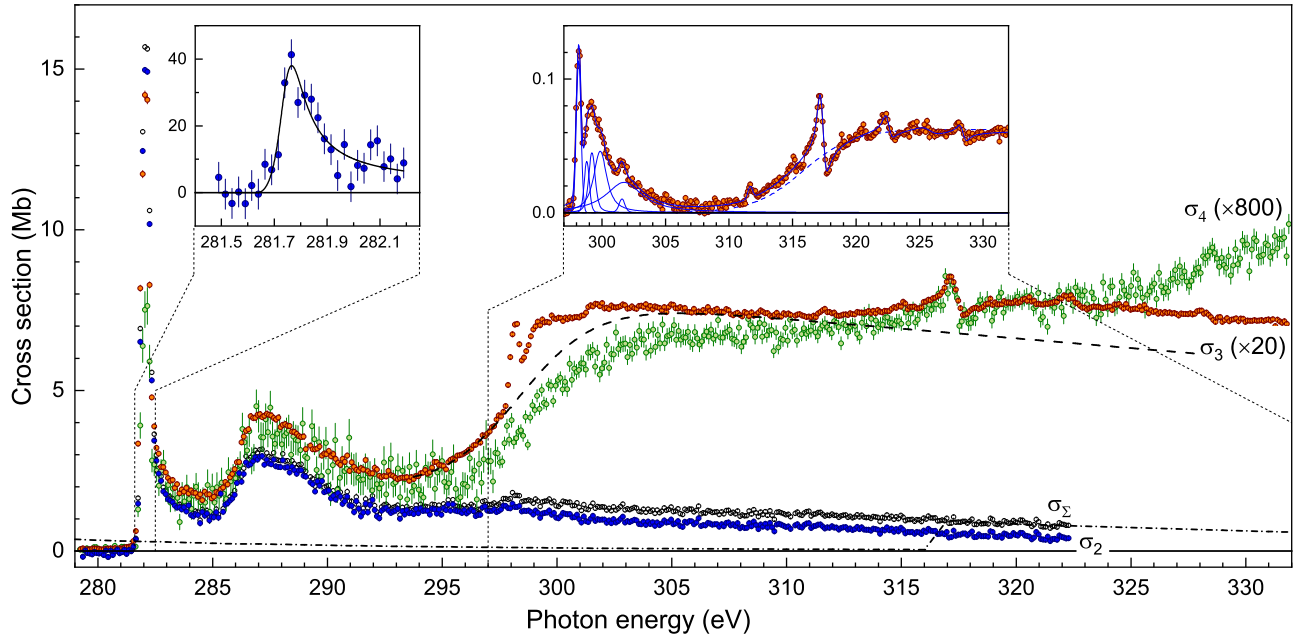}
\caption{\label{fig:Cminus1}Absolute cross sections $\sigma_m$ for $m$-fold detachment of C$^-$ ions \cite{Perry-Sassmannshausen2020}, that were measured with the PIPE setup at the XUV beamline P04 of the PETRA\,III synchrotron light source. The open symbols represent the sum cross section $\sigma_\Sigma = \sigma_2+\sigma_3+\sigma_4$. The black dash-dotted line is the absorption cross section of neutral carbon according to Henke \etal \cite{Henke1993}. The left inset shows the earlier result of Walter et al.~\cite{Walter2006a} for double detachment in the vicinity of the $K$-shell detachment threshold that was obtained at beamline 10.0.1 of the Advanced Light Source (ALS, Berkeley, USA). It should be noted, that the energy scale of the ALS experiment appears to be shifted by -0.25 eV as compared to the energy scale of the PIPE experiment. However, this shift is within the mutual uncertainties of both energy scales. The right inset displays the cross-section difference between the measured cross section $\sigma_3$ for {net} triple detachment of C$^-$ and the dashed line in the main panel of this figure. The thick solid line results from a resonance fit to the experimental data. In the fit, the resonance features were represented by Lorentzian and Fano line shapes convolved with a Gaussian that accounts for the experimental photon energy spread \cite{Schippers2018}. The fitted resonance parameters are provided in Ref.~\cite{Perry-Sassmannshausen2020}. Adapted from \cite{Walter2006a} and \cite{Perry-Sassmannshausen2020} with permission, copyright 2006 and 2020 by the American Physical Society.}
\end{figure*}

Figure~\ref{fig:Cminus1} bears witness to the significant progress that has been achieved at PIPE. The main panel of this figure displays cross sections $\sigma_m$ for net $m$-fold detachment
\begin{equation}\label{eq:reaction}
h\nu + \textrm{C}^- \to  \textrm{C}^{(m-1)+} + m e^-,
\end{equation}
which were measured for $m$=2--5 with low statistical uncertainties over extended energy ranges (for $\sigma_5$ see figure~\ref{fig:Cminus2}). The new data exhibit a wealth of previously unknown thresholds and resonances as compared to the previously available result for double-detachment of C$^-$($1s^2\,2s^2\,2p^3\;^4S$) \cite{Walter2006a}. These previous results are depicted in the left inset of figure~\ref{fig:Cminus1}. Obviously, the data from  PIPE represent a new standard for inner-shell detachment studies with atomic anions.

\subsection{\label{sec:ntres}Near $K$-threshold resonances}

The earlier study by Walter \etal \cite{Walter2006a} focused on the $1s\,2s^2\,2p^4\;^4P$ resonance near the $K$ threshold which was measured with a photon-energy spread of $\sim$120 meV. This is about a factor 4 lower than the 500~meV energy spread of the photon beam used by Perry-Sassmannshausen \etal \cite{Perry-Sassmannshausen2020}. Consequently, the resonance cross section is a factor of $\sim$4 higher in the data of Walter \etal as compared to the double-detachment data of Perry-Sassmannshausen \textit{et~al}. As expected, the integrated resonance strengths of $8.2\pm3.2$~Mb~eV  and of $6.8 \pm 1.0$~Mb~eV, respectively, agree with each other within the rather large uncertainty of the earlier result.

Because of the narrow energy spread in their experiment, Walter \etal \cite{Walter2006a} were able to extract the resonance width of the $1s\,2s^2\,2p^4\;^4P$ resonance from their measurement. To this end, they fitted a Breit-Wigner line shape \cite{Peterson1985} to the experimental data points. This line shape is a function of the photon energy $E$ and is parameterized by the $K$-shell detachment threshold energy $E_\mathrm{th}$, the resonance energy $E_r$, and the resonance width $\Gamma$ as follows:
\begin{equation}\label{eq:BreitWigner}
\sigma(E) = \left\{\begin{array}{lr}\sigma_0 & \textrm{for~} E \leq E_\mathrm{th}\\ \sigma_0 + \frac{\displaystyle A(E-E_\mathrm{th})^{3/2}}{\displaystyle (E-E_\mathrm{r})^2+(\Gamma/2)^2} & \textrm{for~} E > E_\mathrm{th}\\\end{array}\right.
\end{equation}
with $\sigma_0$ and $A$ being an energy-independent background cross section and a scaling constant, respectively. The result of the fit is displayed as a full line in the left inset of figure~\ref{fig:Cminus1}. It agrees very well with the experimental data despite of the fact that Eq.~\ref{eq:BreitWigner} does not account for the experimental energy spread. As already mentioned by Walter \textit{et~al}, the extracted line width $\Gamma =0.11\pm0.04$~eV must therefore be considered as an upper limit. Apart from an insignificant 0.25~eV energy shift (see caption of figure~\ref{fig:Cminus1}), the new double-detachment data from PIPE \cite{Perry-Sassmannshausen2020} confirm the earlier findings of Walter \etal \cite{Walter2006a}.

\begin{figure}
\includegraphics[width=0.95\columnwidth]{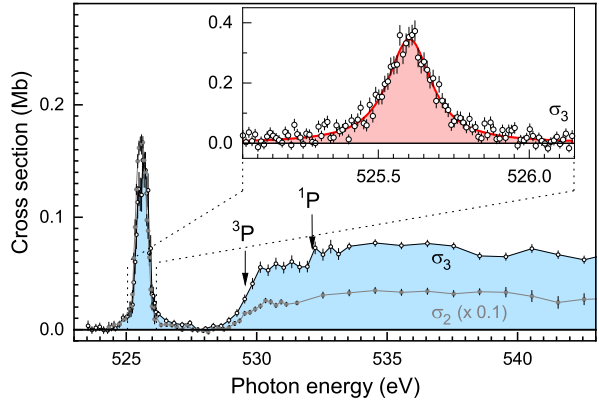}
\caption{\label{fig:Ominus}Measured cross sections $\sigma_2$ (gray-shaded circles) and $\sigma_3$ (open symbols, shaded curves) for double and triple detachment, respectively, of O$^{-}$($1s^2\,2s^2\,2p^5\;^2P$) ions by single-photon impact \cite{Schippers2016a}. The vertical arrows mark the positions of the $1s\,2s^2\,2p^5\;^3P$ and $1s\,2s^2\,2p^5\;^1P$ detachment thresholds at about 529.6 and 532.1~eV, respectively. The resonance at 525.6~eV is associated with a $1s\to 2p$ excitation to the O$^-$($1s\,2s^2\,2p^6\;^2S$) term. The inset shows a Voigt line profile (full line) that has been fitted to high-resolution experimental data (open symbols). The natural line width was determined to be $164\pm14$~meV. Adapted from \cite{Schippers2020}, CC BY licence, copyright 2019 by the authors.}
\end{figure}

The detailed investigation of near $K$-threshold resonances had also motivated the previous inner-shell studies with  Li$^-$($1s^2\,2s^2\;^1S$) \cite{Kjeldsen2001a,Berrah2001}, B$^-$($1s^2\,2s^2\,2p\;^2P$) \cite{Berrah2007a} and  O$^-$($1s^2\,2s^2\,2p^5\;^2P$) \cite{Gibson2012} where double-detachment cross sections were measured over very narrow energy ranges in a similar manner as depicted in the left inset of figure~\ref{fig:Cminus1} for C$^-$ \cite{Walter2006a}. Breit-Wigner line shapes were also found for the near $K$-threshold resonances of  B$^-$($1s^2\,2s^2\,2p\;^2P$) \cite{Berrah2007a}. For Li$^-$ the experimental resolving power was not sufficient for a detailed analysis of resonance line-shapes. The same holds for the O$^-$ data from the ALS \cite{Gibson2012}.

Figure \ref{fig:Ominus} presents the O$^-$ data from PIPE~\cite{Schippers2016a}. As mentioned in section \ref{sec:method}, an ECR ion source was used, although this type of ion source is less suitable for the production of O$^-$ ions than a Cs-sputter source. Nevertheless, absolute cross sections for net double and triple detachment could be measured. In both cross sections, the $1s\,2s^2\,2p^6\;^2S$ resonance at 525.6~eV is prominently visible together with the threshold for direct detachment of a $1s$ electron which sets in at about 529.6~eV. The line-shape of the  $1s\,2s^2\,2p^6\;^2S$ resonance turned out to be purely Lorentzian as revealed by a high-resolution scan (inset of figure~\ref{fig:Ominus}). The experimental photon energy spread  $\Delta E = 40$~meV corresponds to a resolving power $E/\Delta E \approx 13.000$. The natural line-width was obtained by fitting a Voigt line-profile to the measured data points. This  resulted in $\Gamma = 0.164\pm0.014$~eV  or a core-hole lifetime of $4.0\pm0.3$~fs.

Finally, let us note that the experimental photodetachment cross sections of F$^-$($1s^2\,2s^2\,2p^6\;^1S$) at the given experimental sensitivity do not exhibit any resonance structure \cite{Mueller2018b} (figure~\ref{fig:Fminus}). This is attributed to the closed-shell structure of this particular anion.

\subsection{Higher-energy thresholds and resonances}

While all previous $K$-shell detachment experiments with atomic anions were confined to net double detachment in very narrow energy ranges at the $K$-shell detachment threshold (see, e.g., left inset of figure~\ref{fig:Cminus1}), the PIPE data comprise in addition to  double detachment also triple, four-fold, and up to five-fold detachment. These data also extend over much wider ranges of photon energies (Figs.~\ref{fig:Cminus1}, \ref{fig:Fminus}, \ref{fig:Cminus2}). In these extended energy ranges numerous threshold and resonance features have been discovered which had not been discussed in the previously available literature despite of the fact that the atomic systems under investigation are rather fundamental. Of course, this disregard is related to the hitherto restricted availability of corresponding experimental data.

Already the still relatively limited O$^-$ data exhibit an additional threshold feature which is particularly strong in the triple detachment channel (figure~\ref{fig:Ominus}). Apart from the lowest threshold for direct detachment of a $1s$ electron at 529.6~eV a weaker second threshold is visible in the triple-detachment channel at 532.1~eV. These two thresholds are associated with the fine structure of the excited (neutral) atom with a detached $1s$-electron. The respective core-hole terms are $1s\,2s^2\,2p^5\;^3P$ and $1s\,2s^2\,2p^5\;^1P$ as indicated in figure~\ref{fig:Ominus}.

For C$^-$, the thresholds for direct $1s$-detachment are hidden below the strong resonances in the energy range 281--293~eV (figure~\ref{fig:Cminus1}). In fact, quantum mechanical interference between direct and resonant detachment causes the observed asymmetric line shape (Eq.~\ref{eq:BreitWigner}) of the  $1s\,2s^2\,2p^4\;^4P$ resonance discussed above. The second, broad resonance feature at $\sim$287~eV is probably a blend of several individually unresolved resonances associated with $1s\to3p$ excitations as suggested by detailed atomic-structure calculations \cite{Perry-Sassmannshausen2020} (see also section~\ref{sec:theo}). This resonance is visible in all partial cross sections $\sigma_m$ that are displayed in figure~\ref{fig:Cminus1}.

At energies above 293~eV, the cross section for net double detachment of C$^-$ stays nearly constant until 298~eV and from then on decreases monotonically. In contrast, the triple and four-fold detachment cross sections exhibit a pronounced rise that sets in a 295~eV and continues up to 305~eV. This rise is associated with an initial detachment process where one $1s$ electron and one $2p$ electron are removed by a single photon. The subsequent filling of the resulting $K$-shell holes by Auger processes leads to the emission of one or more electrons, such that the initial direct double ionisation event reveals itself only in the net triple and four-fold detachment cross sections and not in the cross section for net double detachment. Direct double core-hole creation will be discussed further in section~\ref{sec:DPI} below.

Above the threshold for $1s+2p$ direct double detachment (and subsequent autoionisation) a number of further resonances appear in particular in the cross section for net triple detachment of C$^-$. These resonances are magnified in the right inset of figure~\ref{fig:Cminus1} which also displays the results of a resonance fit. This fit also suggests another threshold at $\sim$315~eV. At present, a reliable designation of all the fitted features is not at hand, and only the pronounced Fano resonance at 317.3~eV has been tentatively assigned to the $1s\,2s(^3S)\,2p^3(^4P)\,3s(^3P)\,3p\;^4P$ term which can be reached  by a double-excitation process of the C$^-$($1s^2\,2s^2\,2p^3\;^4S$) ground term. \cite{Perry-Sassmannshausen2020}.

In addition to the partial cross sections $\sigma_m$ for $m=2,3,4$, figure~\ref{fig:Cminus1} also displays the sum cross section $\sigma_\Sigma = \sigma_2+\sigma_3+\sigma_4$. It can be argued that, at energies above the $K$-shell detachment threshold, this sum cross section largely corresponds to the C$^-$ absorption cross section. At these energies, net single detachment is negligible since the $K$-hole levels in light ions decay practically exclusively via Auger processes leading to the ejection of further electrons. It should be noted that net single detachment cannot be easily observed by the photon-ion merged-beams technique since this requires the detection of neutral reaction products that propagate along the photon-beam axis. In any case, for energies above the threshold for direct $1s$ ionisation of neutral carbon at $\sim$316.5~eV, $\sigma_\Sigma$ is found to agree excellently with the photoabsorption cross section of neutral atomic carbon \cite{Henke1993} which is also displayed in figure~\ref{fig:Cminus1}. Obviously, the extra electron of C$^-$ does not have a significant influence on the high-energy absorption cross section. At lower energies the absorption is dominated by outer-shell processes that primarily lead to (not measured) single detachment.

\subsection{\label{sec:DPI}Multiple detachment via double core-hole creation}

\begin{figure}[ttt]
\includegraphics[width=0.95\columnwidth]{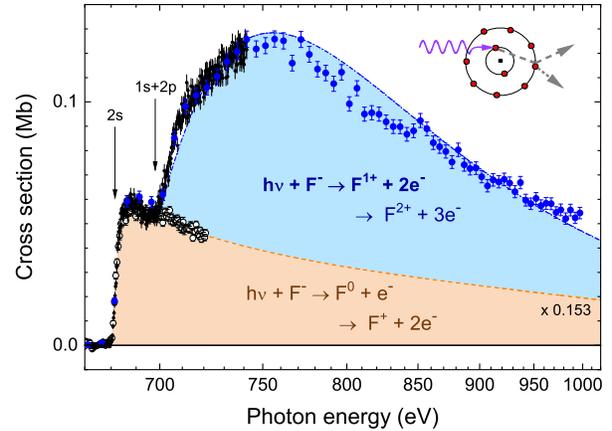}\caption{\label{fig:Fminus}Measured absolute cross sections for double detachment (open symbols, scaled by an overall factor of 0.153) and  triple detachment (small black and large blue full symbols) of  F$^{-}$ ions by single-photon impact \cite{Mueller2018b}. The triple-detachment cross section exhibits a clear signature for a one-photon--two-electron ($1s+2p$) knock-out process $h\nu + \mathrm{F}^- \to \mathrm{F}^+ + 2e^-$. Reproduced from \cite{Schippers2020}, CC BY licence, copyright 2019 by the authors.}
\end{figure}

The fundamental process of direct double photoionisation (DPI), i.e, the removal of two electrons by a single photon, is extremely sensitive to the details of the electron-electron interaction. Therefore, it has been a central topic of atomic physics already for decades (see, e.g., \cite{Wehlitz2010,Sokell2013,Yerokhin2014,Grundmann2018,Chen2020,Andersson2019,Ueda2019} and references therein). Moreover, double core-hole production in atoms and molecules is currently an active topic particularly at VUV and X-ray free-electron lasers \cite{Cryan2010,Young2010,Tamasaku2013,Goldsztejn2016,Eland2010,Berrah2011a,Nakano2013a,Marchenko2017,Feifel2017,Piancastelli2020}.

Inner-shell DPI has been addressed recently in a photodetachment experiment with  F$^-$ anions  \cite{Mueller2018b}. Figure \ref{fig:Fminus} exhibits absolute cross sections for net double and triple detachment of F$^-$. As already mentioned above, these cross sections do not exhibit any  resonances. However, the triple-detachment cross section features a clear signature for the simultaneous knock out of a $1s$ and a $2p$ electron by a single photon. Above a threshold energy of about 700~eV, the associated cross section appears on top of a continuously decreasing \lq{}background\rq\ which is due to single core-hole creation and subsequent emission of two Auger electrons.  This background (dashed curve in figure~\ref{fig:Fminus}) has been modelled on the basis of the separately measured cross section for net double detachment.

The dash-dotted curve in \ref{fig:Fminus} is the sum of the \lq{}background\rq\ and an estimated cross section for $1s+2p$ DPI. The latter has been calculated with the empirical formula developed by Pattard \cite{Pattard2002}, i.e.,
\begin{equation}\label{eq:Pattard}
\sigma(E) = \sigma_M x^\alpha\left( \frac{\alpha+7/2}{x\alpha+7/2}\right)^{\alpha+7/2},\; x = \frac{E-E_\mathrm{th}}{E_M-E_\mathrm{th}},
\end{equation}
which parameterizes the DPI cross section in terms of the Wannier exponent $\alpha$ \cite{Wannier1953} and the cross section maximum $\sigma_M$, which occurs at the energy $E_M$. For $1s+2p$ DPI of F$^-$ these parameters were determined by a fit of Eq.~\ref{eq:Pattard} to the measured cross section (table~\ref{tab:Pattard}).

\begin{table}
	\caption{\label{tab:Pattard}DPI parameters as obtained from fitting Eq.~\ref{eq:Pattard} to the experimental cross sections for net triple photodetachment of F$^-$ \cite{Mueller2018b} (figure~\ref{fig:Fminus}) and for net five-fold photodetachment of C$^-$ \cite{Perry-Sassmannshausen2020} (figure~\ref{fig:Cminus2}). For C$^-$, the Wannier exponent $\alpha$ \cite{Wannier1953} was fixed to the value for the doubly detached intermediate C$^+$ ion.}
\begin{indented}
	\item[]	
    \begin{tabular}{llllll}
           \br
			Ion    & Electrons & $E_\mathrm{th}$ & $E_M$     & $\sigma_M$   & $\alpha$  \\
			       & removed & (eV) & (eV)   &   (kb)  &   \\
			\mr
F$^-$ & $1s+2p$ & 698.7  & 763.1 & 95 & 1.1\\
C$^-$ & $1s+2s$ & 336 & 524 & \phantom{9}1.75 & 1.1269 \\
C$^-$ & $1s+1s$ & 668 & 891 & \phantom{9}3.00 & 1.1269 \\
\br
		\end{tabular}
	\end{indented}
\end{table}

The good  agreement between the experimental cross section and the Pattard formula for DPI supports the interpretation, that the measured triple-detachment cross section to a large part is due to simultaneous removal of a $1s$ and a $2p$  electron by a single photon. The result of the subsequent (Auger) deexcitation cascade is independent of the photon energy and the measured cross section shape is thus determined by the initial DPI process. It should be appreciated, that the $1s+2p$ DPI cross section of F$^-$ could be measured over an extended energy range including the cross-section maximum. This by far exceeds what has been achieved in earlier measurements of photodouble detachment of negative ions, which were restricted to just a few hundred meV near the associated threshold \cite{Donahue1982,Bae1983,Bae1988}.

\begin{figure}
\includegraphics[width=0.95\columnwidth]{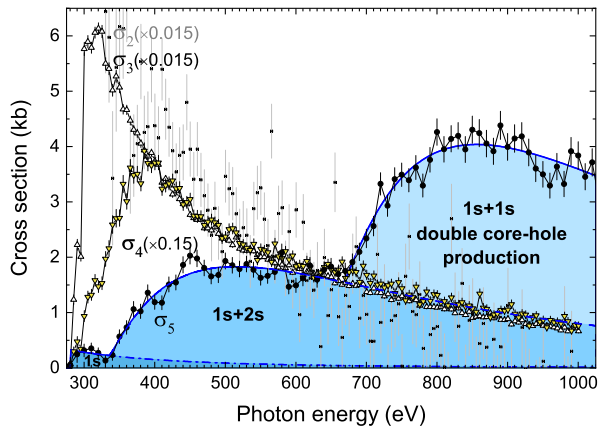}
\caption{\label{fig:Cminus2}Measured cross sections $\sigma_m$ for $m$-fold detachment ($m$=2--5) of C$^-$ ions for energies to beyond the threshold for double K-shell detachment \cite{Perry-Sassmannshausen2020}. The cross section $\sigma_5$ exhibits a clear signature for multiple detachment via double core-hole creation. The shaded curves are fitted model cross sections for single $1s$ photoionisation and for $1s+1s$ DPI. The DPI cross sections were modelled with Eq.~\ref{eq:Pattard} using the parameters from table~\ref{tab:Pattard}. Reprinted with permission, copyright 2020 by the American Physical Society.}
\end{figure}

For C$^-$ the measurements could be extended to even beyond the threshold for double $K$-hole creation \cite{Perry-Sassmannshausen2020} (figure~\ref{fig:Cminus2}). A prominent feature in the cross-section $\sigma_5$ for net five-fold detachment is associated with $1s+1s$ DPI visible at photon energies beyond approximately 700 eV. Moreover, this feature follows the photon-energy dependence of Eq.~\ref{eq:Pattard}. In addition, this cross section exhibits a signature of $1s+2s$ DPI as well. These processes cannot be observed in the other product channels where they are masked by competing processes including the creation of a hole with simultaneous excitation of a valence electron and their subsequent autoionization.

A rough estimate for the ratio of double-to-single $K$-hole production can be inferred from the magnitudes of the cross section for net double and fivefold detachment. This ratio amounts to about 1\% and agrees well with the ratio of  0.3\%, that was measured for $1s$ and $1s+1s$ inner-shell photoionisation of neon, and the theoretical prediction to increase with decreasing nuclear charge \cite{Southworth2003a}. The determination of a more accurate value from the C$^-$ data require a better knowledge of the various deexcitation rates and the formation of the various product charge states, than what is currently at hand. It is interesting to note that the ratio of double vs.\ single core-hole production is much larger in atoms and atomic anions as compared to molecules. For carbon-containing molecules this ratio was found to be typically only 0.1\% \cite{Lablanquie2016}. At present, there is no explanation for this difference.

\section{\label{sec:theo}Theoretical developments}

The experimental findings summarized above challenge  atomic structure theory in several ways.
Difficulties arise especially from (1) the coupling of the inner-shell core holes  to the valence shells, (2) the strong correlations that govern the dynamics of atomic anions and (3) the complexity and depths of the deexcitation cascades.

Gorczyca \cite{Gorczyca2004a} has reviewed early theoretical work on inner-shell photodetachment of Li$^-$, B$^-$, and  C$^-$  focusing particularly on the near $K$-threshold resonances which are discussed in section~\ref{sec:ntres}. An accurate description of these resonances already required to incorporate the post-collision interaction (PCI) into the formalism in order to describe the K-shell detachment. Indeed, PCI (may) result in the re-capture of the initially emitted photoelectron following the Auger decay of the $K$-hole (see also \cite{Lindroth2004a}) and may significantly reduce the theoretical resonance strength for Li$^-$ leading to an excellent agreement with the experimental values. For C$^-$, in contrast, it was found that PCI does not have a significant effect and good agreement with the experiment available at the time of Gorczyca's review was already obtained with more traditional coupled-channel approaches for the calculation of photoabsorption.

For light ions, inner-shell photodetachment primarily leads to the production of singly charged positive ions. For example, the strength of the C$^-$($1s\,2s^2\,2p^4\;^4P$) resonance is dominated by the contribution of the net-double-detachment channel (figure~\ref{fig:Cminus1})  making up for about 90\% of the total absorption. Thus, the contribution of the triple-detachment channel is smaller than the systematic uncertainty of the experimental cross-section scale (cf.\ section~\ref{sec:method}). Therefore, the theoretical cross sections \cite{Gorczyca2004a} for photoabsorption were rightfully expected to compare well with the measured cross section for net double detachment. However, these expectations do not apply for higher-photon energies (figure~\ref{fig:Cminus2}) and for atomic systems with more electrons where broad distributions of product-ion charge states result from inner-shell ionisation.  For such atomic systems, large-scale calculations of Auger cascades are required for predicting the correct charge-state distributions \cite{Schippers2017,Beerwerth2019}.

In the present context, systematically enlarged multiconfiguration Dirac-Fock (MCDF) \textit{ab-initio} calculations  have been performed for resonant inner-shell photodetachment of O$^-$ \cite{Schippers2016a}. These computations showed that the calculated resonance width and branching fractions for double and triple detachment only agree with the experimental findings if double shakeup (and shakedown) of the valence electrons and the rearrangement of the electron density in the course of the autoionisation are taken into account \cite{Fritzsche2012a,Beerwerth2017}. These higher-order multi-electron processes are manifestations of the inner-shell and intra-shell correlation effects that govern the inner-shell photo-detachment dynamics of most atomic anions.

A particular challenge for atomic theory is DPI and, in general, all processes with two electrons in the continuum. These processes rigorously require to construct scattering states with two outgoing (continuum) electrons.  In section \ref{sec:DPI}, therefore, the  experimental cross sections are compared with a semi-empirical scaling formula, since quantum theoretical results are not available. A step forward in this direction has been made very recently by Kheifets \cite{Kheifets2020} who derived the correct shape of the $1s+2p$ DPI cross section for F$^-$ from perturbative and coupled-channel treatments of electron-impact ionisation. The calculations for DPI of C$^-$ \cite{Perry-Sassmannshausen2020} revealed that the double removal of both $K$-shell electrons by a single photon is accompanied, in addition,  by shake processes which make the theoretical treatment even more demanding.

Despite the extraordinary scale of the above mentioned computations significant differences, e.g., between theoretical and experimental absolute cross sections still remain \cite{Berrah2007a,Schippers2016a,Kheifets2020}. For example, a full theoretical description of the rich C$^-$ resonance structure discovered by Perry-Sassmannshausen \etal \cite{Perry-Sassmannshausen2020} (figure~\ref{fig:Cminus1}) is well beyond current reach.  Further efforts and improvements of state-of-the-art atomic theory are therefore required, although some major developments in this direction are already under way \cite{Fritzsche2019}.

\section{\label{sec:conc}Conclusions and Outlook}

Experimental developments in synchrotron-radiation and ion-beam technologies of the past decade have led into a new era of inner-shell photoionisation studies with ionised small quantum systems. As discussed above, the recent experimental results for inner-shell detachment of atomic anions bear vivid witness of this development. The high photon flux at the PETRA\,III beamline P04 as well as the excellent product-ion selectivity of the photon-ion merged-beams end-station PIPE allows one to measure extremely small atomic cross sections in the kilobarn range. This facilitates the observation and quantitative measurement of rare atomic processes such as the simultaneous removal of two inner-shell electrons by a single photon.

The available photon energy range at PETRA\,III beamline P04 permits the investigation of single and double $K$-shell photodetachment of heavier ions up to Cl$^-$ and Na$^-$, respectively. Photodetachment measurements with Si$^-$ have already been performed and are currently being analysed. Further steps towards enlarging the experimental data base on inner-shell processes in atomic anions are planned. This will also stimulate further development of the theoretical methods and might eventually lead to a better understanding of the correlated dynamics in these fundamental atomic systems.

\section*{Acknowledgments}

We thank our collaborators Levente Abrok, Sadia Bari, Alexander Borovik~Jr, Kristof Holste, Ron Phaneuf, Simon Reinwardt, Sandor Ricz, Kaja Schubert, Florian Trinter, and Jens Viefhaus for having actively taken part in beam times at PETRA\,III as well as Randolf Beerwerth and Sebastian Stock for their theoretical calculations. We acknowledge DESY (Hamburg, Germany), a member of the Helmholtz Association HGF, for the provision of experimental facilities. Parts of this research were carried out at PETRA\,III and we would like to thank Kai Bagschik,  Jens Buck, Frank Scholz, J{\"o}rn Seltmann, and Moritz Hoesch for assistance in using beamline P04. We are grateful for support from Bundesministerium f{\"u}r Bildung und Forschung within the \lq{}Verbundforschung\rq\ funding scheme (Grant Nos.\ 05K16GUC, 05K16RG1, 05K16SJA, 05K19GU3, and 05K19RG3) and from Deutsche Forschungsgemeinschaft (DFG, project Schi~378/12-1). MM acknowledges support by DFG through project SFB925/A3.

\section*{References}


\providecommand{\newblock}{}

\end{document}